\documentclass{ article}
\usepackage{graphicx} 
\usepackage{amsmath,amssymb,url}
\usepackage{xcolor}
\usepackage{csvsimple}
\usepackage{graphicx} 
\usepackage{subcaption}     
\usepackage{caption}  

\usepackage{float}

\usepackage{geometry, hyperref}

\usepackage{csvsimple}
\usepackage{siunitx}
\usepackage{tikz}
\usepackage{graphicx}
\usetikzlibrary{shapes.geometric, arrows}

\tikzstyle{imgnode} = [rectangle, rounded corners, minimum width=2cm, minimum height=2cm, draw=black, fill=white]
\tikzstyle{arrow} = [thick,->,>=stealth]

\title{Financial Anomaly Detection for the Canadian Market}
\author{Luigi Caputi and Nicholas Meadows}
\date{March 2026}

\begin{document}

\maketitle
\begin{abstract}
     In this work we evaluate the performance of three classes of methods for detecting financial anomalies: topological data analysis (TDA), principal component analyis (PCA), and Neural Network-based approaches. We apply these methods to the TSX-60 data to identify major financial stress events in the Canadian stock market. We show how neural network-based methods (such as GlocalKD and One-Shot GIN(E)) and TDA methods achieve the strongest performance. The effectiveness of TDA in detecting financial anomalies suggests that global topological properties are meaningful in distinguishing financial stress events. 
\end{abstract}
\section{Introduction}

Stock market crashes, such as the October 1987 market crash, the 2008 financial crash, and more recently the financial crash caused by COVID19 are a source of considerable risk and profit for investors \cite{Sornette}, \cite{Sornette2} \cite{COVID19IMPACT}. Thus, understanding them theoretically and ultimately predicting them is of great importance and a source of theoretical research.  
Authors from a variety of fields have provided a number of theoretical formulations of financial crashes. Amongst other viewpoints, authors have described financial crashes as extreme events (\cite{ExtremeEvents}), changepoints in financial time series (\cite{StatsCrashes}) or as outliers (\cite{Sornette}). It is the latter viewpoint that will inform this work. 

A number of recent works (e.g.~\cite{gidea}, \cite{ChinaTDA}, \cite{IndiaTDA}) have explored the use of Topological Data Analysis (TDA) methods to detect financial crashes. Topological data analysis is a relatively recent field that uses topology and geometry to study the ``shape'' of data. The general summary of the findings of these papers is that stock price data exhibited unusual topological behaviour at times close to extreme financial events, such as the 2008 financial crash and the COVID-19 pandemic. 
In this article, we will use similar methods to those of \cite{gidea} in order to compute topological features associated to stock data on a given day $t$, as opposed to the method based on the Vietoris-Rips complex from \cite{IndiaTDA}, \cite{ChinaTDA}. The method of \cite{gidea} is as follows: given a time series associated to the prices of n stocks, and a fixed window size $W$, compute
 the correlation matrix of the stock prices for days $t, t+1, \cdots t+W$. Then construct the weighted graph whose adjacency matrix is the correlation matrix, and compute the persistent homology of the associated directed flag complex of the graph.  In sum, \cite{gidea} uses (topological) anomalies associated with graphs to predict financial crashes. 

The field of graph neural networks (GNN) offers a different perspective on detecting anomalous graphs (see the surveys \cite{GNN-SURVEY-1} and \cite{GNN-SURVEY-2}). In addition, neural networks have been applied to study financial crashes in \cite{NNCrashDetection}. 

The purpose of this article is to study the effectiveness of graph neural network and TDA-based methods in detecting  financial crashes and other extreme events in the Canadian economy. Specifically, we will apply the preprocessing steps similar to those outlined in \cite{gidea} to the stocks comprising the Canada TSX-60 index to reformulate this as a graph anomaly detection problem, and then apply TDA and GNN based graph anomaly detection methods. The ultimate goal is  to use graph anomaly detection to identify early warning signals of extreme financial events in the Canadian market. The code, developed by the second author, is available on github -- \url{https://github.com/njmead811/Financial-Anomaly-Detection-for-the-Canadian-Market}.

Our analysis  indicates that all methods were capable of detecting major crises such as the 2009 financial crisis, the Greek debt crisis, and COVID-19; however, neural network and TDA methods demonstrated higher precision by also capturing smaller-scale market stress events, such as oil price shocks in 2015–2016. Overall, these findings suggest that methods incorporating global structural information -- particularly neural networks and TDA -- are more effective for financial anomaly detection than approaches relying on raw or linearly transformed features.

\section*{Contributions}

CL: Conceptualisation, Methodology, Investigation, Writing

\noindent NM: Software, Data curation, Formal analysis,  Investigation, Methodology, Visualization, Writing

\section{Methodology}
In our analysis of financial crashes, we consider three primary approaches for detecting anomalies: graph neural networks and topological data analysis (TDA) applied to weighted graphs derived from stock prices, as well as principal component analysis (PCA) applied to the corresponding weighted adjacency matrices. The overall procedure is summarized as follows:

\begin{enumerate}
\item{\textbf{Data:} We use daily price data for the stocks composing the TSX-60 over the period 2005-2021.}
\item{\textbf{Networks construction:} We construct a sequence of weighted graphs by computing correlation matrices of stock log-returns over sliding windows. 
}
\item{\textbf{Graph Neural Networks:} We apply unsupervised graph neural network methods to construct representations of the data. Then compute an anomaly score based on the loss function used to train the neural network.}
\item{\textbf{PCA:} We vectorize each correlation matrix by flattening it and apply principal component analysis for dimensionality reduction. 
}
\item{\textbf{TDA:}: We compute the directed flag complex of each weighted graph and compute the L1 and L2 norms of the resulting persistent barcodes.}
\item{\textbf{Anomaly Detection}: We apply off the-shelf unsupervised anomaly detection methods (Mahalanobis Distance and Local Outlier Factor) to the feature vectors obtained from  PCA and TDA.}
\item{\textbf{Anomaly Scoring:} A graph is classified as anomalous if its score lies above the 97.5th percentile of the empirical distribution. These anomalous graphs are then used to signal potential financial extreme events. }
\end{enumerate}

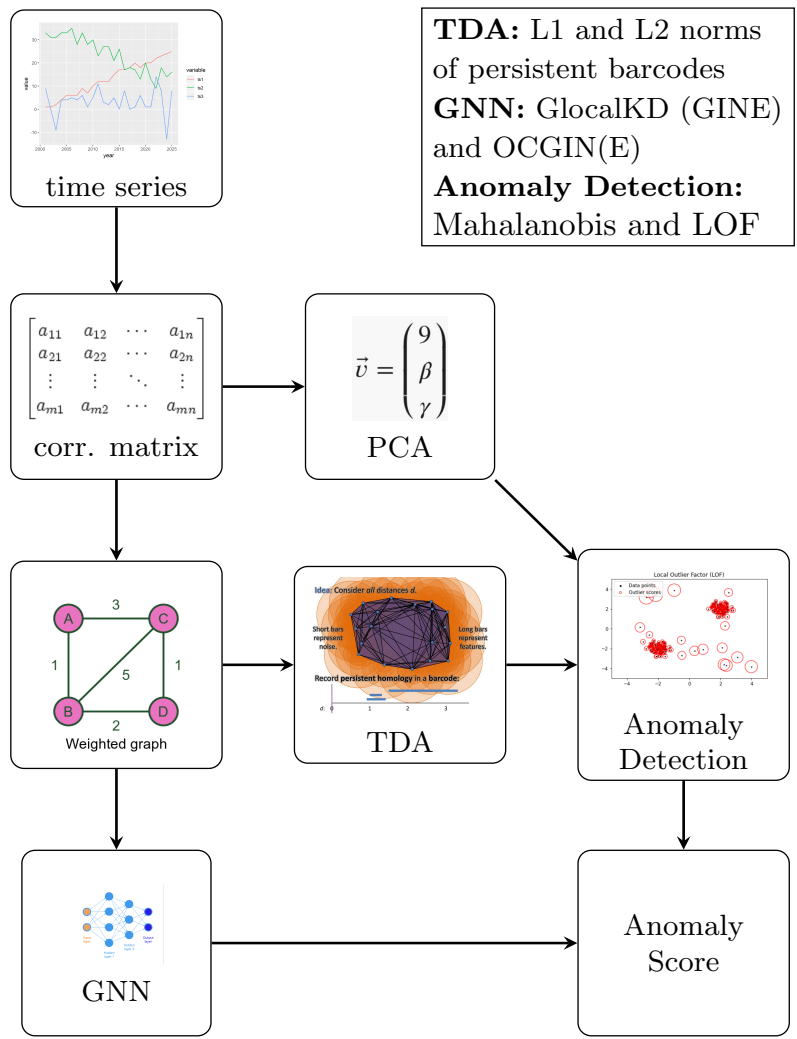
\begin{figure}
    \centering
\resizebox{0.7\textwidth}{0.7\textheight}{
\begin{tikzpicture}[node distance=3cm] 

\node (timeseries) [imgnode] {
  \begin{minipage}{2cm}
    \centering
    \includegraphics[width=2cm]{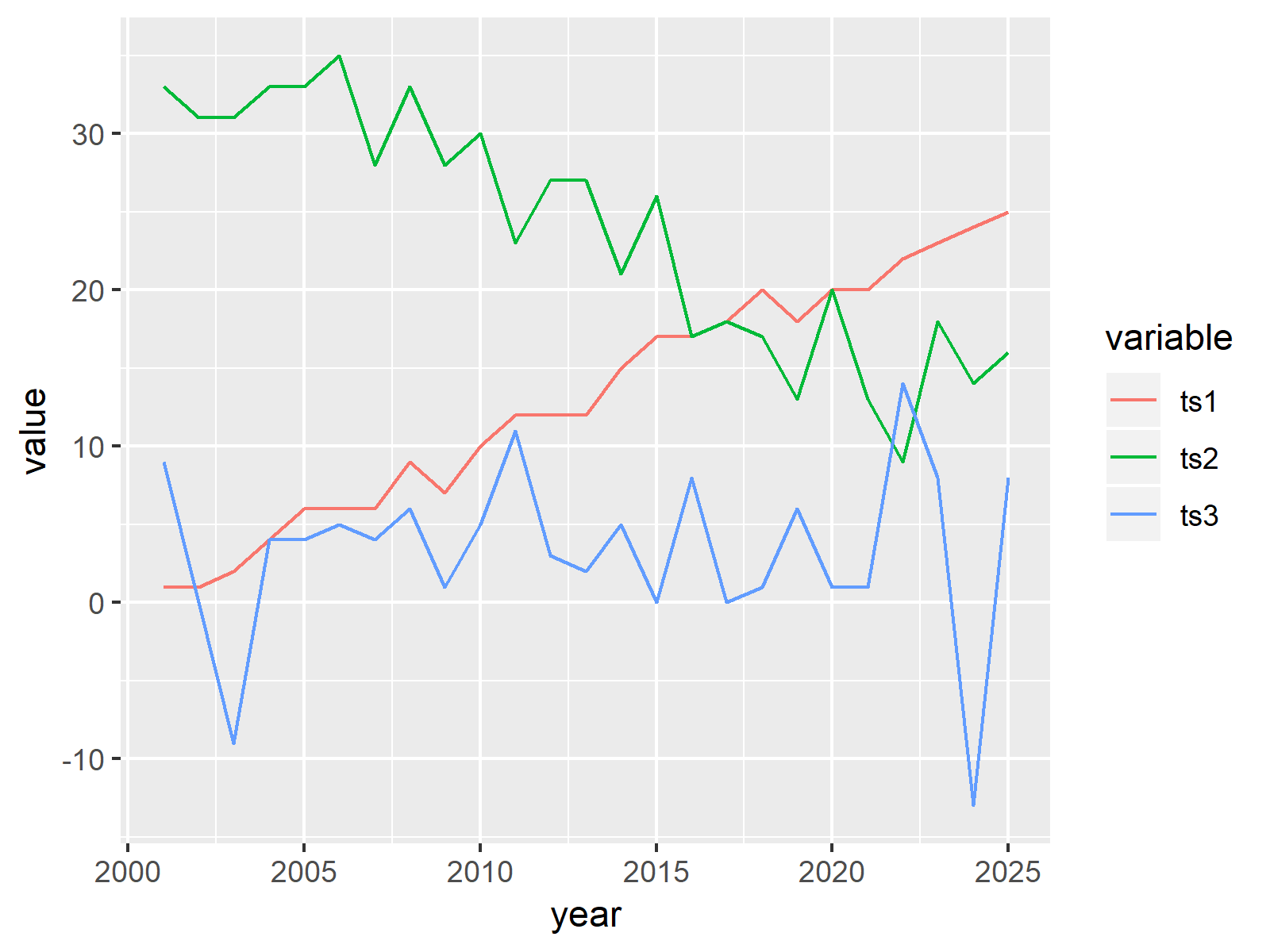}\\
    \small time series
    \end{minipage}
};
\node (correlation) [imgnode, below of=timeseries] {
  \begin{minipage}{2cm}
    \centering
    \includegraphics[width=2cm]{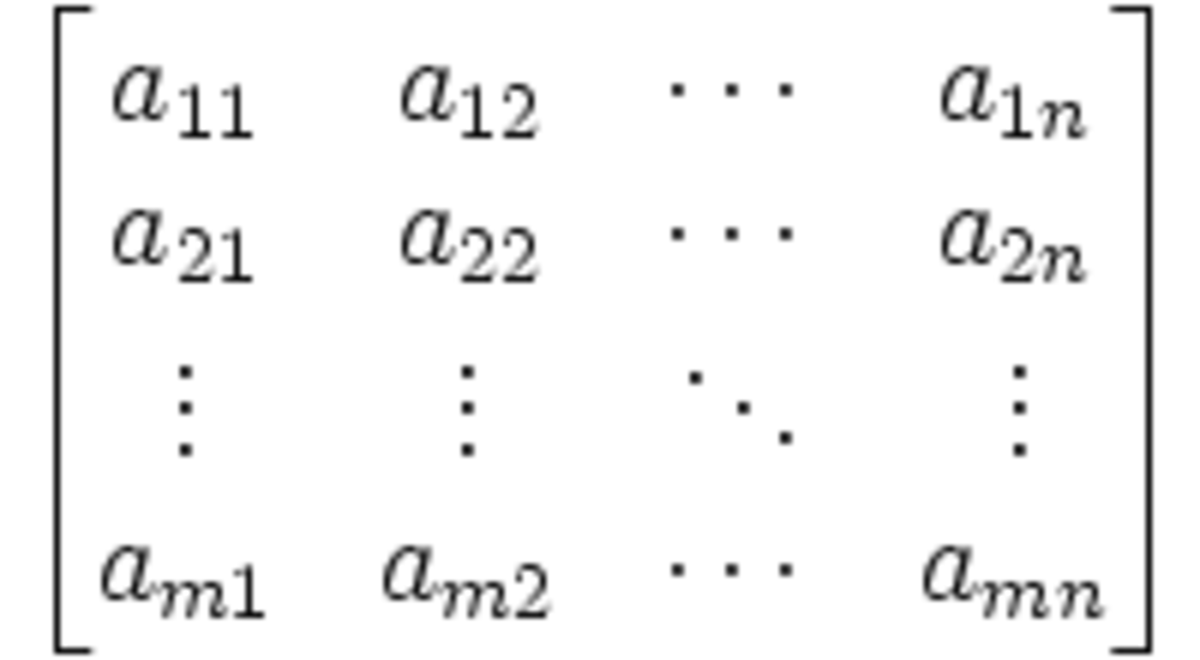}\\
    \small corr. matrix
    \end{minipage}
};
\node (graph) [imgnode, below of=correlation] {\includegraphics[width=2cm]{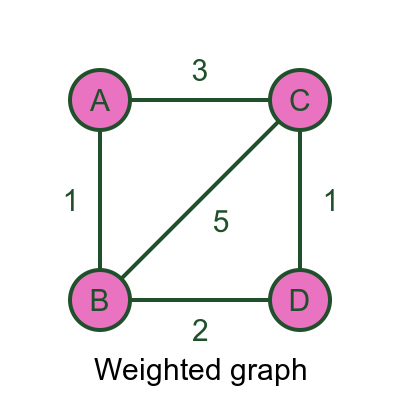}};
\node (GNN) [imgnode, below of=graph] {
 \begin{minipage}{1cm}
    \centering
    \includegraphics[width=1cm]{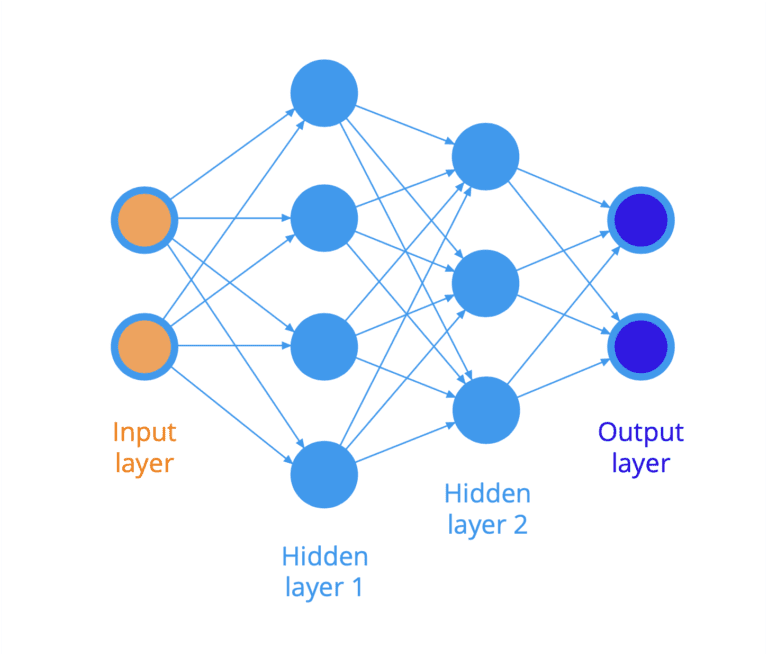}\\
    \small GNN
    \end{minipage}
};

\node (PCA) [imgnode, right of = correlation] {
    \begin{minipage}{1cm}
    \centering
    \includegraphics[width=1cm]{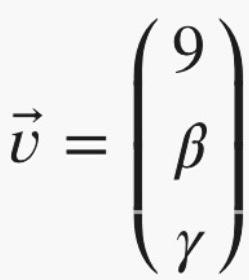}\\
    \small PCA
    \end{minipage}
};

\node (TDA) [imgnode, right of = graph] {
    \begin{minipage}{2cm}
    \centering
    \includegraphics[width=2cm]{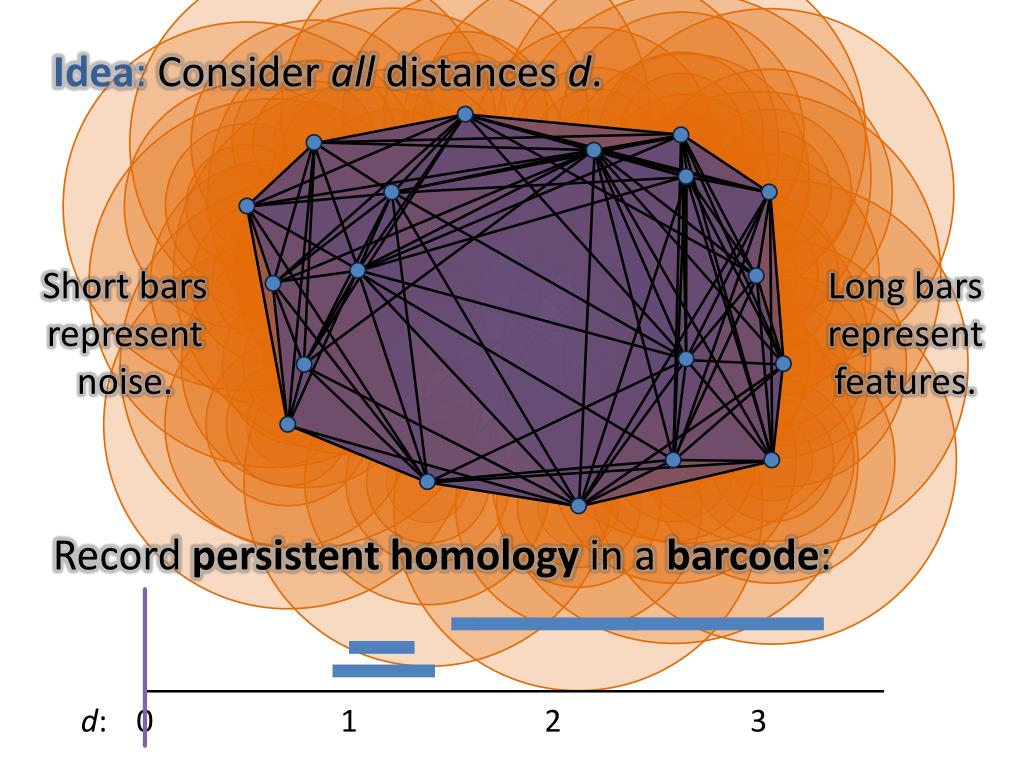}\\
    \small TDA
    \end{minipage}
};

\node (anomaly) [imgnode, right of = TDA] {
    \begin{minipage}{2cm}
    \centering
    \includegraphics[width=2cm]{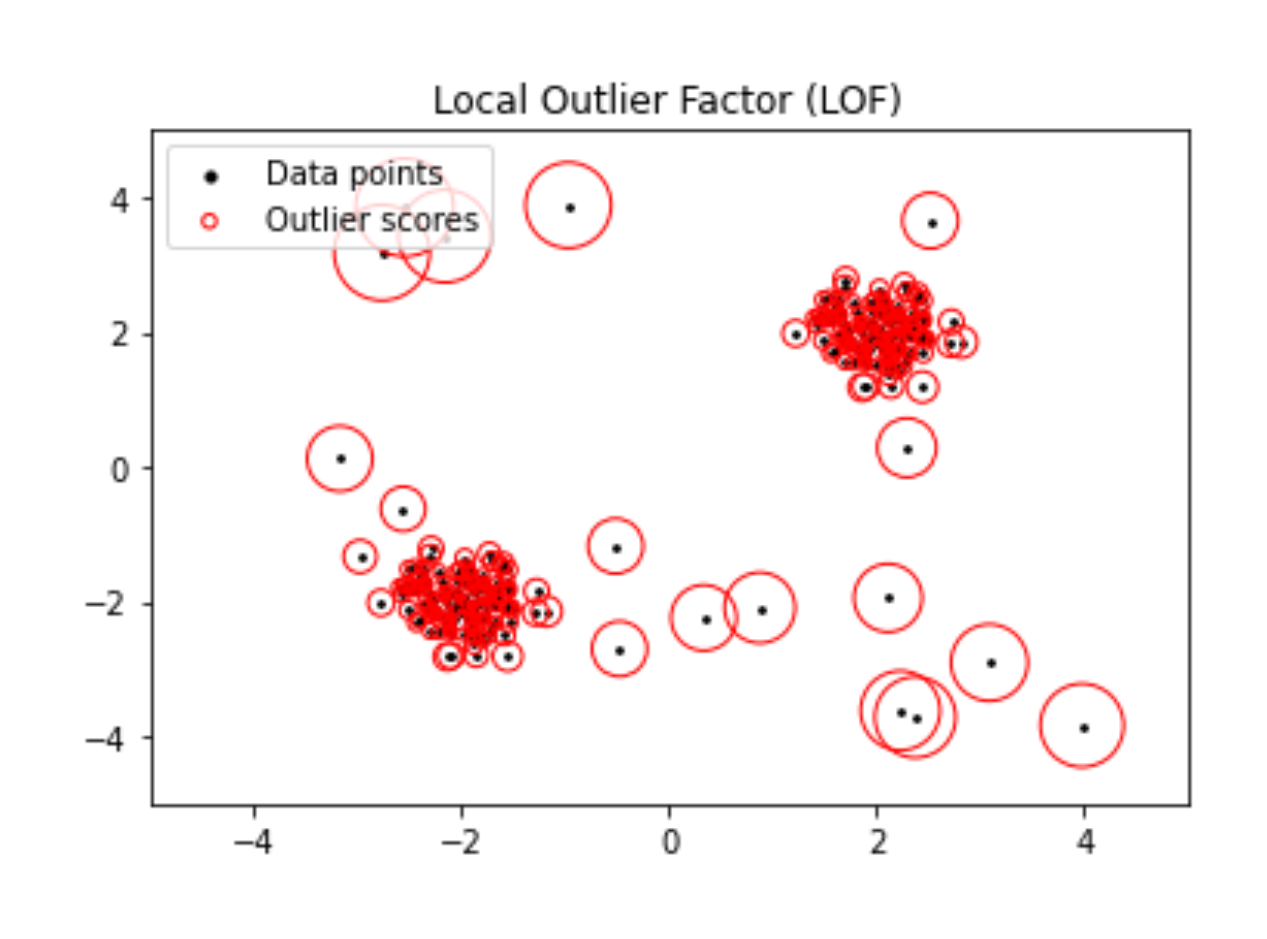}\\
    \small Anomaly Detection
    \end{minipage}
};

\node (score) [imgnode, below of = anomaly] {
    \begin{minipage}{2cm}
    \centering
    \small Anomaly Score
    \end{minipage}
};

\node[draw, align=left] at (5.2, -0.2) {
\small \textbf{TDA:} L1 and L2 norms \\ \small of persistent barcodes \\ 
\small \textbf{GNN:} GlocalKD (GINE) \\ \small and OCGIN(E)
\\ 
\small \textbf{Anomaly Detection:} \\ Mahalanobis and LOF
};

\draw [arrow] (timeseries) -- (correlation);
\draw [arrow] (correlation) -- (graph);
\draw [arrow] (graph) -- (GNN);
\draw [arrow] (correlation) -- (PCA);
\draw [arrow] (graph) -- (TDA);
\draw [arrow] (PCA) -- (anomaly);
\draw [arrow] (TDA) -- (anomaly);
\draw [arrow] (anomaly) -- (score);
\draw [arrow] (GNN) -- (score);

\end{tikzpicture}}
\caption{Pipeline of the Main Analysis}
    \label{fig:pipeline}
\end{figure}

These steps are depicted and summarized in Figure~\ref{fig:pipeline}.
We now proceed providing a more detailed account of these steps. 

\subsection{Data}\label{datasection}
Financial stress is defined as simultaneous financial market turmoil among the most important asset classes. 
In this work we are particularly interested in detecting financial stress events. We focus on the period over the period 2005-2021 for which we have available data. 

We use historical stock prices, downloaded from the Yahoo! Finance service, belonging to the S\&P/TSX 60 Index (TSX-60) of the 60 large companies listed on the Toronto Stock Exchange.  We consider only stocks traded between January 2005 and December 2021. This restriction results in $N=39$ stocks. For daily data, this leads to a time series of length 4254. We used the daily adjusted closing prices. The \emph{logarithmic return } as the first differences of the log-transformed prices: $q^i_t=\log(p^i_t/p^i_{t-1})$, where $p^i_t$ denotes the adjusted daily closing price of stock~$i$ at time~$t$. 

Considering  financial stress events in the Canadian stock market, the major ones are defined  as the seven prominent peaks in the Canadian Financial Stress Index associated to significant financial events recorded in~\cite[Figure 5]{CFSIFoundations} over the period 2005-2021 and  correspond to the following episodes: September 2007 (US mortgage crisis), January 2009 (aftermath of 2008 financial crisis), October 2011 (Greek debt crisis), April 2013 (taper tantrum), January 2015 ( oil prices (Western Canadian Select (WCS)) falling below \$ 40), February 2016 (oil prices (WCS) falling below \$ 20) and April 2020 (COVID19 crisis). We will use these as the main events.

 We note here that the Canadian Financial Stress Index, that we chose as main reference,  is a composite measure of systemic financial market stress; in addition to the equity, government bonds, and foreign exchange  markets, it is considered also  the money market, the bank loans market, the corporate sector, and the housing sector. Further, CFSI captures the co-movement across market segments, which tends to be stronger during systemic events; as a consequence it  ignores the correlation across market segments that occurs during systemic events, leading to  an index that better aligns with known episodes of financial stress in Canada. Last but not least, the housing sector considered in the CFSI is an important source of shocks for the canadian economy. These are the reasons why we considered it as main measure for reference. We wish to point out here also that the CFSI is closely related to international financial stress measures such as the OFR Financial Stress Index or the St.~Louis Fed Financial Stress Index.  However, while international indices capture broad, cross-country or global financial conditions, the CFSI is specifically tailored to the Canadian economy, incorporating domestic variables that reflect Canada’s banking system, interest rate spreads, and exchange rate dynamics.
 
\subsection{Networks construction}\label{sec:construction}
We follow the methods of \cite[Section 3.3.2]{gidea}, to which we refer for more details. We briefly recall the construction of the networks from the financial data described in the previous section. 

For each stock $s_{1}, \cdots, s_{N}$ comprising the TSX-60 (for  $N=39$), we compute the log of the daily closing prices: 
$$
q_{t}^{s_{i}} = \log(p_{t}^{s}/p_{t-1}^{s})
$$
where, as before, we denote by $p_{t}^{s_{i}}$ the closing price of stock $s_{i}$ on day $t$.
Recall taht Convergent Cross Mapping (CCM) is a method for detecting causality in coupled nonlinear dynamical systems, first introduced by Sugihara et al.~\cite{sugihara}. Then, for a fixed window size $W$ (we use $W=25$ days) and each pair of stocks $s_{i}, s_{j}$, we then compute the CCM correlations $c_{i, j}^{t}$ of the time series 
$$
q_{t}^{s_{i}}, q_{t+1}^{s_{i}} \cdots q_{t+W}^{s_{i}}, \, \, \, \, q_{t}^{s_{j}} \cdots q_{t+W}^{s_{j}}
$$
to obtain a matrix $C^{t}$, with $C^t_{i, j} := c_{i, j}$. Following \cite[Section 3,2]{gidea}, we threshold $C^t$ by replacing negative entries with $0$. In turn, we can construct the corresponding weighted graph whose adjacency matrix is the correlation matrix.

\subsection{Graph Neural Networks:} 

An attributed graph consists of a graph $(V, E)$ along with node and edge attributes $X \in \mathbb{R}^{|V| \times m}, Y \in \mathbb{R}^{|E| \times k}$. We denote the attributes of vertex, $v$ and edge $e$ by $X_{v}, Y_{e}$ respectively. 

Mathematically, a graph neural network $NN$ consists of finite series of node embeddings $H_{v}^{i} \in R^{m}, i = 1, \cdots n, v \in V$, which are defined inductively by a formula:

\begin{align*}
H^{0}_{v} = X_{v} \\ 
a_{v}^{i} = \mathbf{AGGREGATE}(H_{w}^{i-1}; \, \,  w \in N(v)) \\
H_{v}^{i} = \mathbf{COMBINE}(H_{v}^{i-1}, a_{v}^{i})
\end{align*}

where $\mathbf{AGGREGATE}, \mathbf{COMBINE}$ are functions that aggregate information from the embeddings of neighboring nodes of $v$ and use this to compute a new node embedding. We think of the final node embedding $H^{n}_{v}$ as the node embedding produced by the graph neural network. Oftentimes, we apply a pooling operation $\mathbf{POOL}$ to the final node embeddings to obtain a graph embedding. We will write 
$$
NN(X, Y, A) := \mathbf{POOL}(H_{v}^{n}; \, \, v \in V), NN(X_{v}, Y, A) := H_{v}^{n},
$$ where $A$ is the adjacency matrix of the graph, X, Y are its node and edge attributes. For more information on the basic theory of graph neural networks consult \cite[Chapter 4]{GraphLearning}.
\\

 The papers \cite{GNN-SURVEY-1} and \cite{GNN-SURVEY-2} provide modern surveys on using graph neural networks for anomaly detection. We now briefly describe the specific GNN-based anomaly detection methods we use in this paper. 

\subsubsection{One-Shot GNN}\label{ocginoverview}
Suppose that we have a family of graphs $\{ G_{1} =(X_1, Y_1, A_1), \cdots ,G_{n} = (X_n, Y_n, A_n) \}$.
In deep one-class learning (\cite{DEEPONESHOT}), for a neural network $NN$, the objective is to minimize 
\begin{equation}\label{OCGINLOSS}
\frac{1}{N} \sum_{i=1}^{n} | NN(X_{i}, Y_i, A_i) - c |^{2} +  \Phi(\Theta)
\end{equation}
where $c$ is some vector and $\Phi(\Theta)$ is some auxiliary task related to the hyperparameters (e.g.~regularization). Essentially the neural networks learns a representation of the graphs centered around~$c$, with anomalous graphs having a larger distance from the center. The regularization term helps to prevent the neural network from learning the constant function $c$ (see the discussion of \cite[Section 3.3]{DEEPONESHOT}). 

In this work we implement a slight variant of the OCGIN architecture from \cite[Section 3.3.2]{ONESHOTGIN}; $c$ is the average of the representation of all the graphs on initialization and $NN$ is the exact architecture from \cite[Section 3.3.2]{ONESHOTGIN}, except that we replace the GIN convolution (\cite[equation 4,1]{POWERFULGNNs}) with a GINE convolution (\cite{GINELAYER}) to handle edge attributes.

\subsubsection{Knowledge Distillation GNN (Glocal KD (GINE))}\label{kdbackground}

In knowledge distillation (see \cite{KD-OVERVIEW}), a simpler student neural network is trained to mimic a larger pretrained network. The idea behind using knowledge distillation to detect graph anomalies is that the student network will be able to capture the majority of graph patterns, and anomalous graphs will have large reconstruction errors (see \cite[V.C]{GNN-SURVEY-2})

In \cite{GlocalKD}, the teacher network is a randomly initialized neural network
$NN(-; \hat{\Theta})$, whose parameters we freeze. We will train another randomly initialized neural network $NN(-;\Theta)$ with the same architecture to mimic the teacher network. Specifically, we choose the loss function
\begin{equation}\label{KDLoss}
\lambda L_{node} + L_{graph}
\end{equation}
where 
$$
L_{node} = \frac{1}{|V|}\sum_{v \in V} |NN(X_{v}, Y, A; \Theta) - NN(X_{v}, Y, A; \hat{\Theta})|^{2}
$$
and 
$$
L_{graph} =  |NN(X, Y, A; \Theta) - NN(X, Y, A; \hat{\Theta})|^{2}
$$
The loss function is chosen under the assumption that an anomalous graph will have anomalous nodes and anomalous global properties. The parameters $\lambda$ will control how much emphasis is given to each. This loss function gives the anomaly score for a graph.

\subsection{Principal Component Analysis} We flatten each $C_{t}$ into a $N \times N $ vector, and apply principal component analysis (PCA) to the resulting vectors to reduce the dimensionality. 

\subsection{Topological Data Analysis}

Topological Data Analysis is a recent and active research field of
applied algebraic topology, whose main goal is to study the structure of datasets by means of topological frameworks. Among the tools
of TDA maybe the most common is nowadays Persistent Homology, which we shall also use here. We now briefly describe our pipeline, referring to~\cite{arXiv:1506.08903} for a comprehensive introduction to the field, or to~\cite{neuro, zbMATH07955445, caputi2025integral} for examples of applications.

For a weighted graph $G$ with $N$ edges, it is customary in topological data analysis to compute persistent homology invariants from the sequence of clique complexes, also known as ``flag complexes'' associated to $G$. We briefly recall the construction. 

We start by filtering a weighted directed graph $G$ by thresholding the weights on the edges: if $w_0 < \dots < w_N$ are the ordered weights of the edges of $G$, we define $G[w_i]$ to be the induced (unweighted) subgraph of $G$ consisting of the same vertices as $G$, and with edges precisely the edges of $G$ of weight  $\leq w_i$. This yields a filtration of directed graphs
\begin{equation}\label{eq:graph_filtration}
G[w_{0}]\rightarrow G[w_{1}] \rightarrow \dots \rightarrow G[w_{N}] \ ,
\end{equation}
that is a sequence of directed graphs and inclusions. 

A directed $n$-clique of directed graph $H$ is a subgraph of $H$ on a collection of vertices $(v_1,\dots, v_n)$ with the property that there is a directed edge $(v_i,v_j)$ if and only if $i<j$. Then,  we define $\mathrm{dFl}(H)$ to be the  simplicial complex on the directed cliques of $H$ and called the \emph{directed flag complex} of~$H$. The filtration of directed graphs \eqref{eq:graph_filtration}
given by the weights, induces  the filtration 
\[
\mathrm{dFl}(G[w_{0}])\rightarrow \mathrm{dFl}(G[w_{1}]) \rightarrow \dots \rightarrow \mathrm{dFl}(G[w_{N}]) 
\]
of directed flag complexes.
The persistent homology groups of the weighted directed graph~$G$ are then defined as the persistent homology groups associated to the filtration of directed flag complexes~\cite{zbMATH01883340}. 

In this work we shall consider persistent homology groups in dimension $0$ and $1$, as main topological indicators. We can visualize the information provided by persistent homology
 by constructing a planar diagram
called a Persistence Diagram, stably with respect to the input data~\cite{zbMATH05126703}. A persistent homology class born at time $b$ and that died at time $d$ is represented in the diagram by the pair $(b,d)$. In this work, we shall measure the distance between diagrams by using the associated $L_1$ and $L_2$ norms~\cite{gidea2018topological}. That is, if $g$ is the dimension of the persistence diagrams and $F_g$ the set of persistent features in the form of pesrsistent diagrams, the associated $L_1$-norm is given by the formula
\[
\sum_{f\in F_g} d_f-b_f
\]
where $d_f$ and $b_f$ are the deaths and births of the feature $f$. Likewise, the $L_2$-norm is given by
\[
 \sqrt{\sum_{f\in F_g}(d_f-b_f)^2}
\]
We could also consider $L_p$-norm for any $p$, but in this work we shall only consider $p=1,2$.

\subsection{Anomaly Detection}\label{anomalydetection}

We shall use two main anomaly scores, the Mahalanobis distance and Local Outlier Factor, both applied to the feature vectors obtained from the PCA and TDA. 

The Mahalanobis distance, was first introduced by Mahalanobis in 1936, and it is a statistical measure recognized for its efficacy in identifying multivariate anomalies.  It has gathered increasing attention also in financial applications  for its potential in outlier detection. In \cite{10.1108/AJAR-09-2018-0032}, the Mahalanobis distance was used  in the task of detecting financial anomalies and assessing creditworthines, revealing its value in the field.  These anomalies, in financial contexts, could be indica-
tive of significant events like fraud or market crashes. Furthermore,  methods like cluster analysis and PCA, in conjunction with the Mahalanobis distance  have seen
widespread application in finance~\cite{demirhan2024financial}.

The Local Outlier Factor (LOF) algorithm is an unsupervised method used to detect anomalies in unlabeled data~\cite{ADESH2024104923}. 
In short, the Local Outlier Factor  and the Mahalanobis Distance are both used for anomaly detection, but they differ in their general approach. If the LOF is a density-based, unsupervised method that evaluates how isolated a data point is by comparing its local density to that of its neighbors, the Mahalanobis Distance is a distance-based measure that determines how far a point lies from the global mean, taking into account feature correlations through the covariance matrix. As a result, it is mainly suited for identifying global outliers and typically assumes the data follows a multivariate normal distribution. On the other hand, LOF is more effective at detecting both local and global outliers in datasets with complex or varying distributions. While LOF is more flexible and better for detecting  local anomalies, Mahalanobis Distance is computationally simpler and works well when the data distribution is well-defined and approximately Gaussian. For these reasons, in this work we make use of both the anomaly detectors.

\subsection{Anomaly Scoring}

We compute the anomaly score for each pair of choice of TDA (e.g. $L^{1}, L^{2}$ norm) and PCA features (e.g. PCA for various choices of dimension) features and anomaly detection method from Section~\ref{anomalydetection}. 

For the neural networks, we assign them an anomaly score based on the reconstruction error of their training objectives. In particular, for the neural network from Section~\ref{ocginoverview}, we use \ref{OCGINLOSS} without the regularization term, and for the neural network of Section~\ref{kdbackground}, we use \ref{KDLoss}. 

In any case, we define a graph to be anomalous for a given anomaly detection method, if the anomaly score is above a certain threshhold. In our course this is $97.5 \%$.

\section{Experiments and Results}

\subsection{Experiment Description}

In this section we evaluate the performance of three classes of methods for detecting financial anomalies (that is, topological data analysis (TDA), principal component analyis (PCA), and Neural Network-based approaches) applied to the TSX-60 data to identify major financial stress events in the Canadian stock market described in \ref{datasection}. 

As described in Section~\ref{sec:construction}, we associate to the index~TSX-60   (a sequence of) weighted graphs constructed from the (CCM) correlations between the time series of log-returns.
Each anomaly detection method  assigns an anomaly score to each business day via the anomaly score of the corresponding graph. We say that the graph is \emph{anomalous} if its anomaly score exceeds the 97.5th percentile of the observed distribution. We then predict the occurrence of  a major financial stress event if an anomalous graph is detected within a 50 business days-window preceding the event.

\subsection{Performance Metrics}

Due to the nature of the task, this constitutes an imbalanced classification problem. We recall that for this type of problems  precision and f-score are more appropriate evaluation measures than accuracy.  
In this setting, the \emph{recall} is defined as the proportion of financial stress events that are successfully signaled by the model; this is measured by the ratio of graphs successfully signaling an event versus the number of graphs. 
An event is considered \emph{successfully signaled} if at least an anomalous graph is detected within the 50 business days preceding the event. 

We define the \emph{precision} as  the ratio of successfully signaled events to the total number of events. 
Finally,  the \emph{f-score} is defined by the classical formula in terms of recall and precision.  

\subsection{Software}

We computed the TDA features using the pyflagser package~\url{https://github.com/giotto-ai/pyflagser}. We used stock price data from the yahoo finance package (yfinance). We implemented Mahalanobis distance using the scipy library, and local outlier factor and PCA using scikit-learn. 

We implemented the neural networks using pytorch-geometric, due to the fact that we somewhat deviated from the original architectures from the literature. In particular, we used their built-in GINE convolution layer (\cite{pytorchgine}). The code for this paper is available on github at \url{https://github.com/njmead811/Financial-Anomaly-Detection-for-the-Canadian-Market}.

\subsection{Hyperparameters for TDA and PCA Methods}

In these methods,  each graph in the sequence   of weighted graphs constructed from the (CCM) correlations associated to the index~TSX-60 is vectorized using PCA or TDA.
Then an anomaly score is assigned via an unsupervised detection method, specifically the Mahalanobis distance or the Local Outlier Factor (LOF).
For the PCA-based analysis, we consider both the raw feature vectors, and the reduced representations of dimension 10 and 100. For the TDA-based analysis we use the persistent features (that is, the $L_1$ and $L_2$ norms of the persistent diagrams in homological dimension $0,1$) derived from $H_{0}$ and $H_{1}$.

The Mahalanobis distance does no require any hyperparameter tuning. For LOF, we adopt the default parameters from sci-kit learn, with the exception of the number of neighbors parameter. In our experiments we vary this parameter over the set $n = 5, 10, 15, 20, 25, 30$.

\subsection{Hyperparameters for Graph Neural Network Methods}

We implement the OCGIN architecture as in \cite[Section 2.2.2]{ONESHOTGIN} including the reccomendedation regularization techniques (see also \cite{DEEPONESHOT}), the only difference being that we replace the GIN layer with the GINE layer \cite{GINELAYER} to deal with edge weights. In particular, we set the bias terms in each fully connected layer to $0$ to avoid the problem of hypersphere collapse (see \cite[Proposition 3.3.2]{DEEPONESHOT}) and utilize weight decay. Our training hyperparameters are as follows: weight decay (0.001, 0.0001, 0.00001, 0.000001), learning rate (0.01, 0.001, 0.0001, 0.00001), batchsize (25, 50, 100), number of layers (2, 3) and hidden dimension (10). 
\\

We implement the architecture from GlocalKD. Although the authors there used a GCN as their base neural network, we have elected to use a GINE neural network instead. We use the following hyperparameters: 
learning rate (0.01, 0.001, 0.0001, 0.00001), batchsize (25, 50, 100), number of layers (2, 3), $\lambda$ (0.1, 0.5, 0.9), hidden dimension (10). 
 
\subsection{Experiment results and discussion}

For the period of 2005-2021, we construct a sequence of weighted graphs from the TSX-60 by applying a sliding window to correlation matrices of stock log-returns.  
We then evaluate the ability of three classes of graph anomaly detection methods—TDA, PCA, and neural network-based approaches—to detect financial extreme events as defined by the CFSI.

\begin{figure}[h!]
\centering
\begin{filecontents*}{output.csv}
Method,Recall,Precision,F1 Score
Glocal KD (GINE),0.56,0.86,0.68
One-Shot GIN,0.47,0.86,0.6
LOF L1PH,0.52,0.57,0.55
LOF L2PH,0.49,0.71,0.58
MAH L1PH,0.5,0.71,0.58
MAH L2PH,0.46,0.71,0.56
LOF RAW,0.27,0.29,0.28
MAH RAW,0.2,0.86,0.32
LOF PCA (dim=10),0.18,0.71,0.28
MAH PCA (dim=10),0.47,0.43,0.45
LOF PCA (dim=10),0.22,0.43,0.29
MAH PCA (dim=100),0.47,0.43,0.45
\end{filecontents*}
\csvautotabular{output.csv}
\centering
\caption{Scores for Different Anomaly Detection Methods (TSX-60)}
    \label{fig:scores1}
\end{figure}

The f-scores for all methods are summarized in Figure~\ref{fig:scores1}. 
Overall, neural network-based methods  (GlocalKD (GINE), One-Shot GIN(E)) achieve the strongest performance, with GlocalKD and One-Shot GINE attaining f-scores of 0.68 and 0.60, respectively. TDA-based methods exhibit intermediate performance, with f-scores in the range 0.55-0.59. The methods based on the raw features, with or without PCA, perform substantially worse (with f-scores 0.28-0.45). 

\begin{figure}[h!]
    \centering
    \begin{subfigure}[t]{0.4\textwidth}
        \centering
        \includegraphics[width=\linewidth]{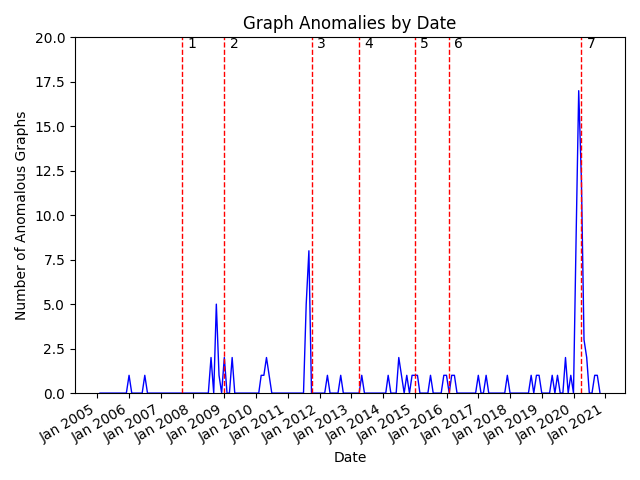}
        \caption{Persistent Barcodes}
        \label{fig:firstbarcodes}
    \end{subfigure}
    \hfill
    \begin{subfigure}[t]{0.4\textwidth}
        \centering
        \includegraphics[width=\linewidth]{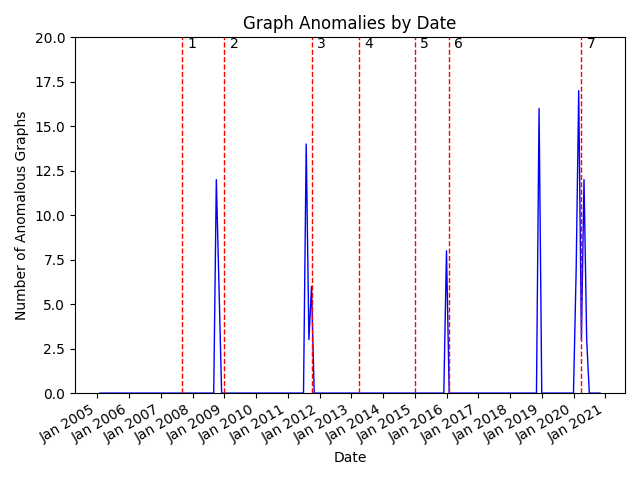}
        \caption{PCA (dim=10)}
        \label{fig:second4b}
    \end{subfigure}

    \vskip\baselineskip

    \begin{subfigure}[t]{0.4\textwidth}
        \centering
        \includegraphics[width=\linewidth]{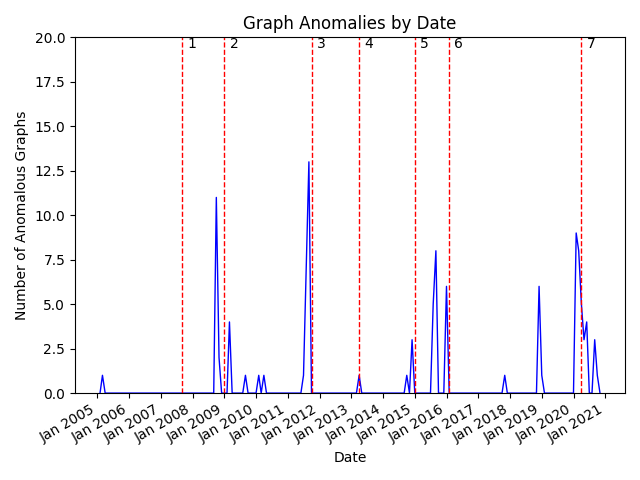}
        \caption{One-Shot GINE}
        \label{fig:firstOneShot}
    \end{subfigure}
    \hfill
    \begin{subfigure}[t]{0.4\textwidth}
        \centering
        \includegraphics[width=\linewidth]{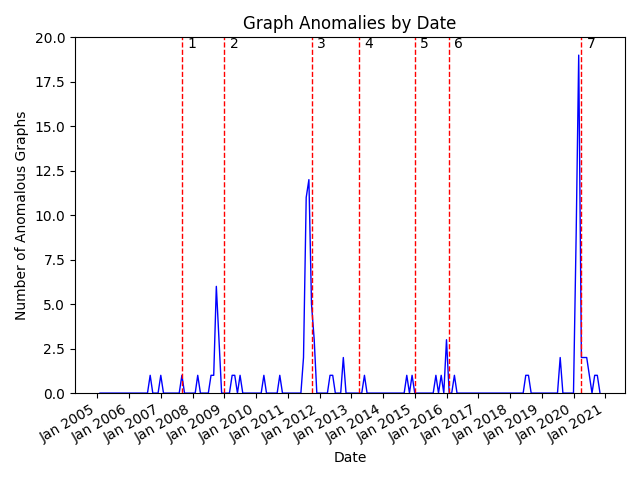}
        \caption{GlocalKD (GINE)}
        \label{fig:second4d}
    \end{subfigure}

       \caption*{\textbf{Legend:} 1-Start of Mortgage Crisis, 2-Peak of 2008  Crisis, 3-Greek Debt Crisis, 4- Taper Tanrum, 5 - Oil $< \$40$, 6 - Oil $< \$20$, 7- COVID19}
     \caption{Anomalies Detected By Different Methods (TSX-60)}
    \label{fig:multi-pane}
\end{figure}

Further insight on the experiment is summarised in Figure~\ref{fig:multi-pane}, where we recorded the number of anomalous graphs detected per month for the two neural network methods, as well as the best-performing TDA and PCA-based methods. We observe that all methods are capable of identifying major financial crises, such as the 2009 financial crisis, the Greek debt crisis, and the COVID-19 crisis. However, there is  a large difference in the best recall between the PCA-methods  and both the TDA and neural network approaches; with the latter ones achieving approximately  10 percent higher recall.
Moreover, neural networks and TDA-based methods achieved notably higher precision, as opposed to PCA. This is because they were able to detect not only major events, but also smaller-scale periods of financial stress. For instance, the TDA-based model identifies a large number of small spikes  in 2015 and 2016, a period associated with declining oli prices and near-recession conditions in the  Canadian economy. Similarly the OCGine had a small spike in early 2015 followed by a larger spike in early 2016 corresponding to historical lows in the oil prices.

We now briefly comment on the selection of the methods. First, we point out here that the preprocessing method from \cite{gidea} (as opposed to other TDA-based methods) was chosen so that we could use the general framework of graph anomaly detection to study the problem of detecting financial crises.  
In choosing the neural network-based methods, we based our work on~\cite{GNN-SURVEY-1} and~\cite{GNN-SURVEY-2}. In particular,  \cite[Table VII]{GNN-SURVEY-2} gives a comprehensive list of state of the art graph-level anomaly detection methods. Many of these methods were not suitable as they were either difficult to implement or designed for specialised tasks, like explainable AI or heterogenous graphs. 
Of the remaining options discussed in~\cite{GNN-SURVEY-2}, three of them were examples of one-shot graph learning. We chose the OCGin(E) from these, due to the well-known connection between GIN and the Wiesfaler-Lehman isomorphism test~\cite{POWERFULGNNs}. 
We also chose the Glocal (\cite{GlocalKD}) method due to its relative ease of implementation and its flexibility in that the authors mention that it can be used with a wide variety of base graph neural networks (\cite[4.1.1]{GlocalKD}). Once again our hypothesis of the importance of global properties led us to choose the GINE neural network as our basis due to the importance of global properties and edge weights. 

Our choice of hyperparameters (eg.~batch size and weight decay) follow  standard practice. For GlocalKD (GINE), we select a relatively small value of the parameter~$\lambda$, reflecting our emphasis on global graph anomalies  over local node-level deviations.

From our results, of interest to the authors is that  the effectiveness of TDA in detecting financial anomalies suggests that global properties (related to homotopy/isomorphism type) are of importance in distinguishing financial stress events. This is a result which does depend on the networks at hand. For similar based analysis in neuroimage, TDA-based approaches have not consistently outperformed simpler correlation-based methods~\cite{neuro}.

Finally, to assess the robustness of our findings,  we replicate the experiments on the Dow-Jones Index. In this case, as shown in the appendix, TDA and Neural Network based methods performed substantially better than PCA.

\appendix

\section{Results for DJIA}

In this section, we repeat the experiments of the preceding section for the Dow Jones Industrial Average (DJIA) of the US stock market. We consider only stocks traded between January 2005 and December 2021, which results in $N=26$ stocks and a time series length of 4254. We consider peaks in the average monthly values of the OFR Financial Stress index (\cite{OFR}), which correspond to the following major events: September 2007 (US mortgage crisis), November 2008 (2008 Financial Crisis), June 2010 (2010 flash crash), October 2011 (Greek Debt Crisis), September 2015 (Stock Market Selloff), February 2018 (early 2018 market correction), March 2020 (COVID19). We record the results in  \ref{fig4} and \ref{fig:multi-pane3} below. As before, neural network methods and TDA based methods performed substantially better than PCA.

\begin{figure}[h]
\centering
\begin{filecontents*}{resultsUS_25_.csv}
Method,Recall,Precision,F1 Score
Glocal KD (GINE),0.74,1,0.85
One-Shot GIN,0.76,0.86,0.8
LOF L1PH,0.36,1,0.53
LOF L2PH,0.33,1,0.49
MAH L1PH,0.64,0.86,0.74
MAH L2PH,0.53,0.86,0.66
LOF RAW,0.43,0.71,0.54
MAH RAW,0.09,0.43,0.15
LOF PCA (dim=10),0.19,0.86,0.31
MAH PCA (dim=10),0.36,0.71,0.47
LOF PCA (dim=10),0.28,0.57,0.38
MAH PCA (dim=100),0.36,0.71,0.47
\end{filecontents*}
\csvautotabular{resultsUS_25_.csv}
\centering
\caption{Scores for Different Anomaly Detection Methods (DJIA)}
\label{fig4}
\end{figure}
\newpage
\begin{figure}[H]
    \centering
    \begin{subfigure}[t]{0.4\textwidth}
        \centering
        \includegraphics[width=\linewidth]{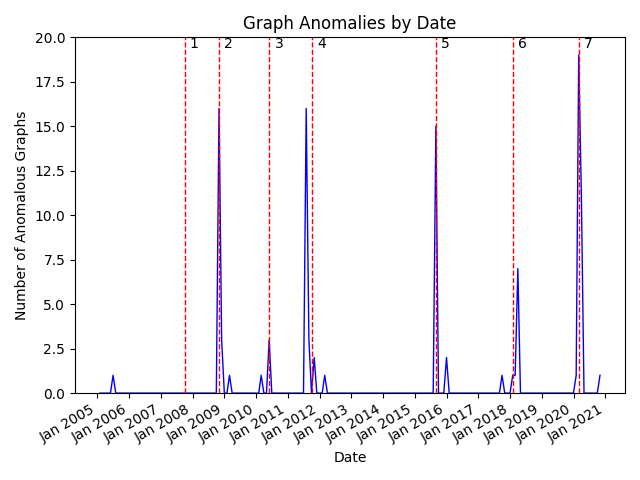}
        \caption{Persistent Barcodes}
        \label{fig:first5bars}
    \end{subfigure}
    \hfill
    \begin{subfigure}[t]{0.4\textwidth}
        \centering
        \includegraphics[width=\linewidth]{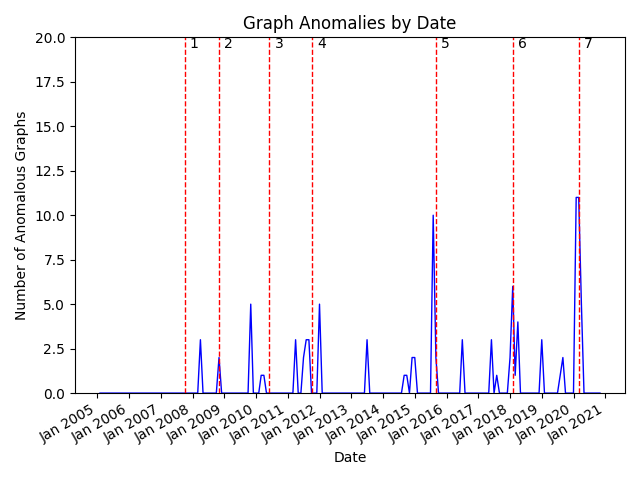}
        \caption{LOF RAW}
        \label{fig:second5b}
    \end{subfigure}

    \vskip\baselineskip

    \begin{subfigure}[t]{0.4\textwidth}
        \centering
        \includegraphics[width=\linewidth]{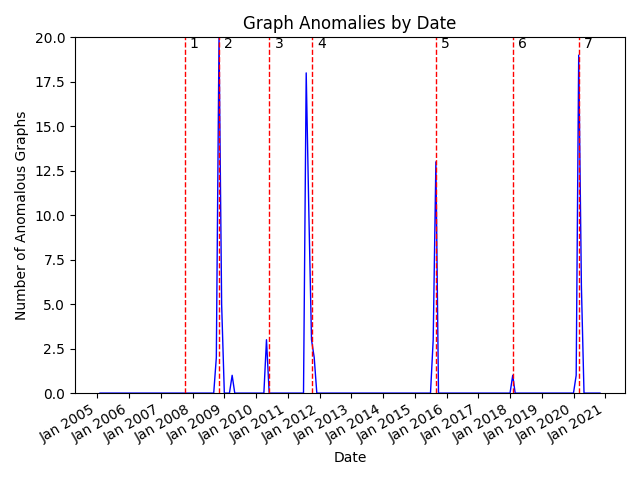}
        \caption{One-Shot GIN(E)}
        \label{fig:first5Oneshot}
    \end{subfigure}
    \hfill
    \begin{subfigure}[t]{0.4\textwidth}
        \centering
        \includegraphics[width=\linewidth]{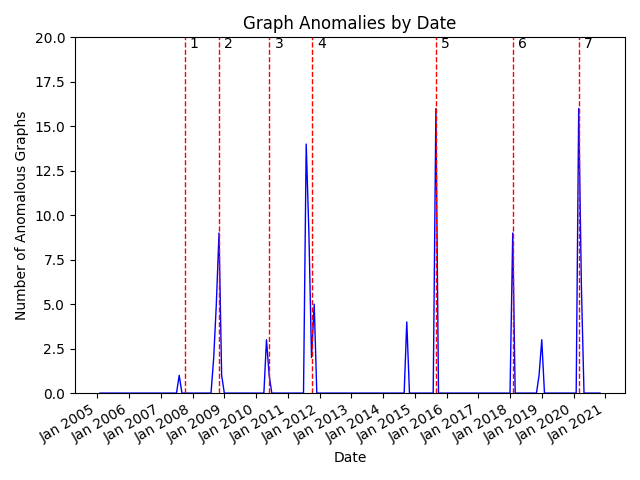}
        \caption{GlocalKD (GINE)}
        \label{fig:second5d}
    \end{subfigure}
     \caption*{\textbf{Legend:} 1 - Start of Mortgage Crisis, 2 -  2008 Financial Crisis , 3 - Flash Crash, 4 - Greek Debt Crisis, 5 - 2015 Stock Market Selloff, 6 - 2018 Stock Market Correction, 7 - COVID19}
    \caption{Anomalies Detected By Different Methods (DJIA)}
    \label{fig:multi-pane3}
\end{figure}

\bibliography{biblio.bib}

\newcommand{\etalchar}[1]{$^{#1}$}
\begin{thebibliography}{MWX{\etalchar{+}}21}

\bibitem[AGSX24]{ADESH2024104923}
A.~Adesh, Shobha G, J.~Shetty, and L.~Xu.
\newblock Local outlier factor for anomaly detection in hpcc systems.
\newblock {\em Journal of Parallel and Distributed Computing}, 192:104923,
  2024.

\bibitem[BZR{\etalchar{+}}22]{KD-OVERVIEW}
L.~Beyer, X.~Zhai, A.~Royer, L.~Markeeva, R.~Anil, and A.~Kolesnikov.
\newblock Knowledge distillation: A good teacher is patient and consistent.
\newblock In {\em 2022 IEEE/CVF Conference on Computer Vision and Pattern
  Recognition (CVPR)}, pages 10915--10924, 2022.

\bibitem[CPH21]{neuro}
L.~Caputi, A.~Pidnebesna, and J.~Hlinka.
\newblock Promises and pitfalls of topological data analysis for brain
  connectivity analysis.
\newblock {\em NeuroImage}, 238:118245, 2021.

\bibitem[CPH25]{caputi2025integral}
L.~Caputi, A.~Pidnebesna, and J.~Hlinka.
\newblock Integral betti signatures of brain, climate and financial networks
  compared to hyperbolic, euclidean and spherical models.
\newblock {\em Scientific Reports}, 2025.

\bibitem[CR24]{zbMATH07955445}
L.~Caputi and H.~Riihim{\"a}ki.
\newblock Hochschild homology, and a persistent approach via connectivity
  digraphs.
\newblock {\em J. Appl. Comput. Topol.}, 8(5):1121--1170, 2024.

\bibitem[CSEH07]{zbMATH05126703}
D.~Cohen-Steiner, H.~Edelsbrunner, and J.~Harer.
\newblock Stability of persistence diagrams.
\newblock {\em Discrete Comput. Geom.}, 37(1):103--120, 2007.

\bibitem[Dem24]{demirhan2024financial}
H.~Demirhan.
\newblock Financial anomalies and creditworthiness: a python-driven machine
  learning approach using mahalanobis distance for ise-listed companies in the
  production and manufacturing sector.
\newblock {\em Journal of Financial Risk Management}, 13(1):1--41, 2024.

\bibitem[Dup20]{CFSIFoundations}
T.~Duprey.
\newblock Canadian financial stress and macroeconomic condition.
\newblock {\em Canadian Public Policy}, 46, 2020.

\bibitem[ELZ02]{zbMATH01883340}
H.~Edelsbrunner, D.~Letscher, and A.~Zomorodian.
\newblock Topological persistence and simplification.
\newblock {\em Discrete Comput. Geom.}, 28(4):511--533, 2002.

\bibitem[Gid17]{gidea}
M.~Gidea.
\newblock Topology data analysis of critical transitions in financial networks.
\newblock In {\em 3rd International Winter School and Conference on Network
  Science}, 2017.

\bibitem[GK18]{gidea2018topological}
M.~Gidea and Y.~Katz.
\newblock Topological data analysis of financial time series: Landscapes of
  crashes.
\newblock {\em Physica A: Statistical mechanics and its applications},
  491:820--834, 2018.

\bibitem[Hu19]{GINELAYER}
W.~et.~al Hu.
\newblock Strategies for training graph neural networks, 2019.
\newblock Accepted as a Spotlight at ICLR 2020 available at
  \url{https://arxiv.org/abs/1905.12265}.

\bibitem[LS24]{NNCrashDetection}
H.~Li and S.~Song.
\newblock Early warning signals for stock market crashes: empirical and
  analytical insights utilizing nonlinear methods.
\newblock {\em EPJ Data Science}, 13, 2024.

\bibitem[LTV19]{10.1108/AJAR-09-2018-0032}
M.~Lokanan, V.~Tran, and Nam~H. Vuong.
\newblock Detecting anomalies in financial statements using machine learning
  algorithm: The case of vietnamese listed firms.
\newblock {\em Asian Journal of Accounting Research}, 4(2):181--201, 08 2019.

\bibitem[LW24]{GraphLearning}
L.~Zhao L.~Wu, P. Cui J.~Pei.
\newblock {\em Graph Neural Networks: Foundations, Frontiers and Applications}.
\newblock Springer Nature, Singapore, 2024.

\bibitem[MF23]{COVID19IMPACT}
R.~Mckibben and R.~Fernando.
\newblock The global economic impacts of the covid-19 pandemic.
\newblock {\em Economic Modelling}, 129, 2023.

\bibitem[MP22]{GlocalKD}
R.~Ma and G.~Pang.
\newblock Deep graph-level anomaly detection by glocal knowledge distillation.
\newblock In {\em Proceedings of the International Conference on Learning
  Representations (ICLR)}, 2022.
\newblock Available at \url{https://openreview.net/pdf?id=ryGs6iA5Km}.

\bibitem[MWX{\etalchar{+}}21]{GNN-SURVEY-1}
X.~Ma, J.~Wu, S.~Xue, J.~Yang, C.~Zhou, Q.~Z. Sheng, H.~Xiong, and L.~Akoglu.
\newblock A comprehensive survey on graph anomaly detection using deep
  learning.
\newblock {\em IEEE Transactions on Knowledge and Data Engineering},
  35(12):12012 -- 12038, 2021.

\bibitem[OFR]{OFR}
Ofr financial stress index.
\newblock \url{https://www.financialresearch.gov/financial-stress-index/}.

\bibitem[OPT{\etalchar{+}}15]{arXiv:1506.08903}
N.~Otter, M.~A. Porter, U.~Tillmann, P.~Grindrod, and H.~A. Harrington.
\newblock A roadmap for the computation of persistent homology.
\newblock Preprint, {arXiv}:1506.08903 [math.{AT}] (2015), 2015.

\bibitem[pyt]{pytorchgine}
Pytorch geometric: Gineconv layer.
\newblock
  \url{https://pytorch-geometric.readthedocs.io/en/latest/generated/torch_geometric.nn.conv.GINEConv.html}.

\bibitem[QTA{\etalchar{+}}25]{GNN-SURVEY-2}
H.~Qiao, H.~Tong, B.~An, I.~King, C.~Aggarwal, and G.~Pang.
\newblock Deep graph anomaly detection: A survey and new perspectives.
\newblock {\em IEEE Transactions on Knowledge and Data Engineering}, 2025.

\bibitem[RMNP22]{StatsCrashes}
A.~Rai, A.~Mahata, M.~Nurujjaman, and O.~Prakash.
\newblock Statistical properties of the aftershocks of stock market crashes
  revisited: Analysis based on the 1987 crash, financial-crisis-2008 and
  covid-19 pandemic.
\newblock {\em International Journal of Modern Physics C}, 33, 2022.

\bibitem[RSL{\etalchar{+}}24]{IndiaTDA}
A.~Rai, B.~Sharma, S.~Luwang, M.~Nurujjaman, and S.~Majhi.
\newblock Identifying extreme events in the stock market: A topological data
  analysis.
\newblock {\em Chaos}, 34, 2024.

\bibitem[RVG{\etalchar{+}}18]{DEEPONESHOT}
L.~Ruff, R.~Vandermeulen, N.~Goernitz, L.~Deecke, S.~A. Siddiqui, A.~Binder,
  E.~Müller, and MariusKloft.
\newblock Deep one-class classification.
\newblock In {\em Proceedings of the International Conference on Machine
  Learning (ICML)}, pages 4493--4402, 2018.
\newblock Available at \url{https://proceedings.mlr.press/v80/ruff18a.html}.

\bibitem[SA06]{ExtremeEvents}
H.~Kantz S.~Albeverio, V.~Jentsch.
\newblock {\em Extreme Events in Nature and Society}.
\newblock The Frontiers Collection. Springer Nature, New York, 2006.

\bibitem[SMY{\etalchar{+}}12]{sugihara}
G.~Sugihara, R.~May, H.~Ye, C.~Hsieh, E.~Deyle, M.~Fogarty, and S.~Munch.
\newblock Detecting causality in complex ecosystems.
\newblock {\em Science}, 338(6106):496--500, 2012.

\bibitem[Sor03]{Sornette}
D.~Sornette.
\newblock Critical market crashes.
\newblock {\em Physics Reports}, 378:1--98, 2003.

\bibitem[Sor17]{Sornette2}
D.~Sornette.
\newblock {\em Why Stock Markets Crash: Critical Events in Complex Financial
  Systems}.
\newblock Princeton Science Library. Princeton University Press, Princeton,
  2017.

\bibitem[XLHJ19]{POWERFULGNNs}
K.~Xu, J.~Leskovec, W.~Hu, and S.~Jegelka.
\newblock How powerful are graph neural networks.
\newblock In {\em Proceedings of the International Conference on Learning
  Representations (ICLR)}, 2019.
\newblock Available at \url{https://openreview.net/pdf?id=ryGs6iA5Km}.

\bibitem[YLW{\etalchar{+}}25]{ChinaTDA}
J.~Yao, J.~Li, J.~Wu, M.~Yang, and X.~Wang.
\newblock Change point detection in financial market using topological data
  analysis.
\newblock {\em Systems}, 13, 2025.

\bibitem[ZL23]{ONESHOTGIN}
L.~Zhao and Akoglu L.
\newblock “on using classification datasets to evaluate graph outlier
  detection: Peculiar observations and new insights.
\newblock {\em Big Data}, 11, 2023.

\end{thebibliography}
\bibliographystyle{alpha}

\end{document}